\documentclass[prc,aps,preprint,showpacs,nofootinbib,showkeys,eqsecnum,tightenlines]{revtex4}    
\usepackage{epsfig}
\def\be{\begin{equation}}
\def\ee{\end{equation}}
\def\bea{\begin{eqnarray}}
\def\eea{\end{eqnarray}}
\def\ot{(1\,\leftrightarrow\,2)}
\def\loh{$\Lambda_{1/2}\,\,$}
\def\bloh{$\bar{\Lambda}_{1/2}\,\,$}
\def\hdr{$\hat d^R$\,\,}

\def\s2q2{(\vec \sigma_2 \cdot \vec q_2)}

\newcommand{\noi}{\noindent}

\def\Journal#1#2#3#4{{#1}{\bf #2} (#4) #3 }
\def\NPA{{ Nucl. Phys.} \bf A}
\def\NPB{{ Nucl. Phys.} \bf B}
\def\PRL{ Phys. Rev. Lett.\,\,}
\def\PRA{{ Phys. Rev.}  A}
\def\PRC{{ Phys. Rev.}  C}

\def\PRW{ Phys. Rev.\,\,}
\def\FBS{ Few--Body Systems\,\,}
\def\PLB{{ Phys. Lett.} \bf B}

\def\EPJA{{ Eur. Phys. J.}\bf A}

\def\PR{ Phys. Rep.\,\,}

\def\IJMPE{{ Int. J. Mod. Phys.} \bf E}

\def\HI{ Hyperfine Interactions\,\,}
\def\SJPN{ Sov. J. Part. Nucl.\,\,}

\def\RMP{ Rev. Mod. Phys.\,\,}
\def\ANP{ Adv. Nucl. Phys.\,\,}
\def\JPG{ J. Phys. \bf G: Nucl.Part. Phys.\,\,}
\def\AJ{ Astrophys. J.\,\,}
\def\CJPB{{ Czech. J. Phys.} \bf B}
\def\PPNP{Prog. Part. Nucl. Phys.\,\,}

\begin{document}
\draft
\title{
{\large{\bf Calculation of Doublet Capture Rate for Muon Capture in
Deuterium within Chiral Effective Field Theory}}}
\author{J.~Adam, Jr.}\affiliation{Institute of Nuclear Physics ASCR, CZ--250 68
\v{R}e\v{z}, Czech Republic }
\author{M.~Tater}\affiliation{Institute of Nuclear Physics ASCR, CZ--250 68
\v{R}e\v{z}, Czech Republic }
\author{E.~Truhl\'{\i}k}
\affiliation{Institute of Nuclear Physics ASCR, CZ--250 68
\v{R}e\v{z}, Czech Republic }
\author{E. Epelbaum}
\affiliation{Institut fuer Theoretische Physik II, Fakultaet fuer
Physik und Astronomie, Ruhr-Universitaet Bochum, 44780 Bochum,
Germany}
\author{R. Machleidt}
\affiliation{Department of Physics, University of Idaho, Moscow, ID
83844-0903, USA}
\author{P. Ricci}
\affiliation{Istituto Nazionale di Fisica Nucleare, Sezione di
Firenze, I-50019 Sesto Fiorentino (Firenze), Italy }

\begin{abstract}

The doublet capture rate \loh  of the negative muon capture in
deuterium is calculated employing the nuclear wave functions
generated from accurate nucleon-nucleon (NN) potentials constructed
at next-to-next-to-next-to-leading order of heavy-baryon chiral
perturbation theory and the weak meson exchange current operator
derived within the same formalism. All but one of the low-energy
constants that enter the calculation were fixed from pion-nucleon
and nucleon-nucleon scattering data. The low-energy constant \hdr
(c$_D$), which cannot be determined from the purely two-nucleon
data, was extracted recently from the triton $\beta$-decay and the
binding energies of the three-nucleon systems. The calculated values
of \loh show a rather large spread for the used values of the \hdr.
Precise measurement of \loh in the future will not only help to
constrain the value of \hdr, but also provide a highly nontrivial
test of the nuclear chiral EFT framework. Besides, the precise
knowledge of the constant \hdr will allow for consistent
calculations of other two-nucleon weak processes, such as
proton-proton fusion and solar neutrino scattering on deuterons,
which are important for astrophysics.

\end{abstract}

\noi 

\noi \hskip 1.9cm 

\maketitle

\section{Introduction}
\label{intro}

The weak nuclear interaction plays crucial role in the formation of
stars in our Universe: it starts the pp chain of the solar burning.
In this chain, the following reactions occur, triggered by the weak
nuclear interaction \cite{AEA}, \bea
p\,+\,p\,&\rightarrow&\,d\,+\,e^+\,+\,\nu_e\,,  \label{pp} \\
p\,+\,p\,+\,e^{-}\,&\rightarrow&\,d\,+\,\nu_e\,,  \label{pep}\\
p\,+\,^{3}He\,&\rightarrow&\,^{4}He\,+\,e^+\,+\,\nu_e\,,  \label{hep}\\
^{7}Be\,+\,e^-\,&\rightarrow&\,^{7}Li\,+\,\nu_e\,,   \label{7bee} \\
^{8}B\,&\rightarrow&\,^{8}Be^*\,+\,e^+\,+\,\nu_e\,.   \label{8b}
\eea

The neutrinos produced in these reactions are messengers from the
very core of the Sun, where the hydrogen burning occurs. Therefore,
their study can provide a valuable information on star formation.
The neutrinos, released in reaction (\ref{8b}) have a continuous
spectrum with the maximum energy 15 MeV and have recently been
registered in the SNO detector \cite{SNOI,SNOII,SNOIII} via the
reactions \bea
\nu_x\,+\,d\,&\rightarrow&\,\nu^\prime_x\,+\,n\,+\,p\,,  \label{NC} \\
\nu_e\,+\,d\,&\rightarrow&\,e^-\,+\,p\,+\,p\,,  \label{CC} \eea also
induced by the weak nuclear interaction.  As a result, the
registered neutrino flux for the neutral current reaction (\ref{NC})
confirmed the validity of the Standard Solar Model. Simultaneously,
the neutral current to charged current ratio \cite{SNOII}
established unambiguously the presence of an active neutrino flavor
other than $\nu_e$ in the observed solar neutrino flux, thus
confirming definitely the phenomenon of the neutrino oscillations
and that the neutrinos possess a finite mass.

It is clear that the precise description of  reactions
(\ref{pp})-(\ref{CC}) is of fundamental value. However, the
reactions (\ref{pp})-(\ref{CC}) cannot be studied experimentally at
present with desired accuracy in terrestrial conditions. In order to
perceive them, one should address other weak processes in
few-nucleon systems that are feasible in laboratories, such as \bea
^{3}H\,&\rightarrow&\,^{3}He\,+\,e^-\,+\,\bar{\nu}_e\,,   \label{3h} \\
\mu^-\,+\,^{3}He\,&\rightarrow&\,^{3}H\,+\,\nu_\mu\,,  \label{mu3hec} \\
\mu^-\,+\,d\,&\rightarrow&\,n\,+\,n\,+\nu_\mu\,.   \label{mudc} \eea

Then, relying on the cosmological principle, one can apply the
acquired knowledge on the weak nuclear interaction also to other
weak processes, occurring in the extraterrestrial conditions. The
reactions (\ref{3h}) and (\ref{mu3hec}) have already been studied
experimentally in great detail. The half-life of the triton is known
with an accuracy $\sim$ 0.3 \%, $(fT_{1/2})_t= (1129.6\pm 3)$ s
\cite{AM}, and the capture rate of muon on $^{3}$He (\ref{mu3hec}),
$\Lambda_0$= 1496 $\pm$ 4 sec$^{-1}$ \cite{AAV,PA} is also known
with the same accuracy. The situation with the reaction (\ref{mudc})
is less favorable so far. Indeed, the last measurements of the
doublet capture rate provided \loh= 470 $\pm$ 29 sec$^{-1}$ \cite{M}
and \loh= 409 $\pm$ 40 sec$^{-1}$ \cite{CEAL}. This unfavorable
situation is expected to change soon in view of a precision
experiment planned by the MuSun Collaboration \cite{MSE}, whose goal
is to measure \loh with an accuracy of $\sim$ 1.5 \%. Keeping in
mind that the two-nucleon system is much simpler than the
three-nucleon one such a precise knowledge of \loh will help to
clarify best the situation in the theory of reactions triggered by
weak nuclear interaction in few-nucleon systems.

The main theoretical problem concerning the above quoted nuclear
reactions is caused by the non-perturbative nature of quantum
chromodynamics (QCD)  at low energies. The way of handling this
obstacle has already been outlined 50 years ago by realizing that
the description of the electro-weak interaction with a nuclear
system, containing nucleons and pions, should be based on the
spontaneously broken global chiral symmetry SU(2)$_L \times$
SU(2)$_R$, reflected in the QCD Lagrangian \cite{AD,AFFR}, see also
\cite{DGH,WQFTII}. Later on, a concept of hidden local symmetry
allowed one to extend the nuclear system to contain also heavy
mesons \cite{HLSLM,BKY}. The nonlinear chiral Lagrangians
\cite{WNLL,OZ,ITI,STG} served then as a starting point for
constructing the one-boson exchange currents in the tree
approximation. The reactions (\ref{NC})-(\ref{mudc}) were studied
with such currents in
Refs.\,\cite{MRTI,MRTV,CiT,CT,ATCS,AHHST,TKK,RTMS}. Calculations
\cite{STG,MRTI,MRTV,ATCS,RTMS} were performed with the nuclear wave
functions derived from one-boson exchange potentials
\cite{BOBEPS,SKTS}, thus providing rather consistent results. We
refer to this concept as Tree Approximation Approach (TAA).  In the
more general Standard Nuclear Physics Approach (SNPA)
\cite{NSAPMGK}, one employs realistic nucleon-nucleon (NN)
potentials which usually  have no relation to applied meson-exchange
currents thus lacking the consistency. Generally, these concepts
describe well the nuclear phenomena triggered by the electro-weak
interaction up to energies $\sim$ 1 GeV
\cite{KTR,ITPR,JFM,DOR,EW,CSc,DFM,SCK,GIG}.

The crucial step allowing one to go beyond the tree approximation
was made by Weinberg \cite{W}. In this work, Weinberg formulated the
principles of chiral perturbation theory ($\chi$PT) which is an
effective field theory (EFT) of QCD at low energies. The effective
chiral Lagrangian is constructed by including all possible
interactions between pions and nucleons consistent with the
symmetries of QCD and, especially, the spontaneously broken
approximate chiral symmetry. Weinberg also established counting
rules \cite{W1,W2} allowing one to classify the contribution of
various terms of perturbative expansion in positive powers of
$q/\Lambda_\chi$, where $q$ is a momentum or energy scale of the
order of the pion mass characterizing a given hadronic system which
is small in comparison with the chiral symmetry breaking scale
$\Lambda_\chi\,\sim$ 1 GeV.

In order to calculate reliably the capture rates and cross sections
of reactions one needs to know accurately the current operators and
the nuclear wave functions (potentials). The weak axial currents
were studied within the EFT in Refs.\,\cite{PKMR,DG}. At leading-
(LO) and the next-to-leading (NLO) orders, the weak nuclear current
consists of the well-known single nucleon terms \cite{MCP,CMS}. The
space component of the weak axial meson-exchange current (MEC),
which contributes to observables in all weak reactions mentioned
above and is, therefore, of the main interest here, appears first at
N$^{3}$LO.  It involves one unknown constant \hdr, \be \hat
d^R\,=\,\hat d_1\,+\,2\hat d_2+\,\frac{1}{3}\,(\hat c_3\, +\,2\hat
c_4)\,+\,\frac{1}{6}\,, \label{reldrcd} \ee where the dimensionless
constants $\hat c_i$ and $\hat d_j$ are given by \be \hat
c_i\,=\,M_N\,c_i\,,\quad\,i=1, \ldots ,4\,,\quad \hat d_j\,=\,
-\frac{M_N f^2_\pi}{g_A}\,d_j\,,\quad j=1,2\,, \label{cidj} \ee with
$c_i$ and $d_j$ being the low-energy constants (LECs) entering the
chiral Lagrangian of the $\pi NN$ system at the NLO \cite{PKMR,DG}.
Further, $M_N$=0.939 GeV is the nucleon mass,  $g_A$ is the nucleon
axial vector coupling constant and $f_\pi$ denotes the pion decay
constant. The linear combination
$\hat d_1\,+\,2\hat d_2$ is often replaced by an effective LEC $c_D$
\cite{EEA}, \be \hat d_1\,+\,2\hat d_2\,=\,-\frac{M_N}{\Lambda_\chi
g_A}c_D\,,\label{CED} \ee where $\Lambda_\chi = 0.7$ GeV.

Let us note that the weak axial currents \cite{PKMR,DG} were derived
without referring to any equation describing nuclear states. The
resolution of the problem of double counting in conjunction with the
Schroedinger equation \cite{MRT} provided a weak axial pion
potential current, the presence of which ensures that the weak axial
MECs satisfy the nuclear PCAC constraint,

\be q_\mu j^a_{\,5\mu}(2,\vec
q)\,=\,\,+\,([\,V_\pi\,,\,\rho^a_{\,5}(1,\vec q)\,]+\ot) +\,if_\pi
m^2_\pi \Delta^\pi_F(q^2)\,M^a_\pi(2,\vec q)\,. \label{NCEt} \ee

Here $V_\pi$ is the one-pion exchange potential,
$\rho^a_{\,5}(1)$ is the one-nucleon axial charge density,
$m_\pi$ is the pion mass,
$\Delta^\pi_F(q^2)$ is the pion propagator and
$M^a_\pi(2)$ is the two-nucleon pion absorption/production amplitude.
The weak axial potential current is not a part of the weak axial currents
of Refs.\,\cite{PKMR,DG}.

Eq.\,(\ref{NCEt}) is a direct analogue of the nuclear conserved vector
current constraint valid for the weak vector MECs,

\be q_\mu j^a_\mu(2,\vec q)\,=\,([\,V_\pi\,,\,\rho^a(1,\vec
q)\,]+\ot)\,. \label{NCVC} \ee

The consistency of calculations requires that both potentials and
currents are derived from the same Lagrangian. Presently,  only few
calculations of weak reactions performed within the EFT approach
fulfill this requirement. Instead, one often adopts the so-called
``hybrid'' approach in which the current operator is constructed
from the EFT as outlined above but the wave functions are generated
either from the one-boson-exchange potentials of the TAA
\cite{SKTS,CDB} or from purely phenomenological potentials
\cite{WSS}. Notice that all these potentials describe low-energy NN
scattering data with $\chi^2$/data $\approx 1$. Besides, when
calculating observables for the weak processes in the three- and
more nucleon systems (see below), also the three-nucleon forces
\cite{CEA,PEA} were addressed. So far, almost all calculations
aiming to study the weak interaction in few-nucleon systems made use
of the precisely known  half-life of the triton to extract \hdr. In
this way, this constant  was determined  in Ref.\,\cite{ASP} from
the reduced Gamow-Teller (GT) matrix element for the reaction
(\ref{3h}), and then the spectroscopic factors $S_{pp}
(0)=3.94\times(1\pm 0.004)\times 10 ^{-25}$ MeV b and
$S_{hep}(0)=(8.6\pm 1.3) \times 10 ^{-30}$ keV b were calculated for
reactions (\ref{pp}) and (\ref{hep}), respectively. Using the same
values of \hdr, the doublet capture rate \loh$=386$ sec$^{-1}$  for
the reaction (\ref{mudc}) was obtained in Ref.\,\cite{APKM} and the
cross sections of the reactions (\ref{NC}) and (\ref{CC}) were
calculated in Ref.\,\cite{ASPFK}. Let us note that the authors
\cite{ASP} employed purely phenomenological 2N AV18 \cite{WSS} and
3N Urbana-IX \cite{PEA} potentials. Similarly, in Ref.~\cite{GAZ},
the capture rate $\Lambda_0 = 1499 \pm 16$ sec$^{-1}$ for the
reaction (\ref{mu3hec}) was obtained based on the same potentials.
In Ref.~\cite{PI1}, Marcucci and Piarulli employed AV18 NN potential
to generate the nuclear wave functions for the process
$^{2}$H($\mu^-,\nu_\mu$)nn and AV18 + Urbana IX 3N  potentials for
the reaction $^{3}$He($\mu^-,\nu_\mu$)$^{3}$H. The weak current was
taken from the $\chi$PT approach with the weak axial potential
current of Ref.\,\cite{MRT} added. The calculations  resulted in
$\Lambda_{1/2}$ = 393.1 $\pm$ 0.8 s$^{-1}$ and $\Lambda_0$ = 1488
$\pm$ 9 s$^{-1}$.

The problem with the above mentioned hybrid calculations is that
they do not allow for a simultaneous determination of the LECs $c_D$
and $c_E$ which govern the short-range part of the 3NF at N$^2$LO.
More precisely, the constant $c_D$ enters not only the weak axial
MEC, but also the contact and one-pion exchange part of the
three-nucleon force whereas the constant c$_E$ controls the strength
of the three-nucleon contact term \cite{UVC,EEA}. These LECs can be
determined by purely hadronic few-nucleon observables such as
e.g.~the triton binding energy and the nucleon-deuteron doublet
scattering length \cite{EEA} or the spectra  of light nuclei
\cite{Navratil:2007we}. Strong correlations between various
few-nucleon observables such as e.g. the Philips and Tjon lines,
however, often prevent from a precise determination of these LECs.
On the other hand, a recent determination of these LECs from the
triton binding energy and the half life \cite{GQN}  seems to be more
robust in this sense.

We also mention a recent work  \cite{RTMS}, where the values of \loh
for the reaction (\ref{mudc}) were  calculated within the TAA. The
nuclear wave functions were derived from the \mbox{Nijmegen I} and
\mbox{Nijmegen 93} potentials \cite{SKTS} and the exchange current
models were constructed within the approach of hidden local symmetry
\cite{STG}. Taking these capture rates as input, then the values of
\hdr were extracted in the hybrid approach, in which the nuclear
wave functions were kept the same as in TAA and the weak axial MECs
were taken from the $\chi$PT approach \cite{PKMR}.

The reactions of muon capture in deuterium and in $^{3}$He have
recently been studied by the Pisa group \cite{PI2} in the SNPA- and
hybrid approaches and within the $\chi$PT as well. The resulting
values of the capture rates are, \mbox{$\Lambda_{1/2}$ = (389.7 -
394.3) s$^{-1}$} and \mbox{$\Lambda_0$ = (1471 - 1497) s$^{-1}$.}
Subsequent calculations by this group \cite{PI3} within the $\chi$PT
provided \mbox{\loh = 400$\pm$3 s$^{-1}$} and \mbox{$\Lambda_0$ =
1494$\pm$21 s$^{-1}$.} We will make a more detailed comparison with
these results in the next section.

Recently, accurate NN potentials at N$^3$LO by Epelbaum, Gl\"ockle
and Mei\ss ner (EGM) \cite{EGM} and Entem and Machleidt (EM)
\cite{EM} have become available. In both cases, the parameters
entering the calculations are standardly extracted from the fit to
the NN scattering data and the deuteron properties, or taken from
the analysis of the $\pi N$ scattering.

The EGM  and EM  potentials differ from each other in several
aspects. First, EGM adopted the so-called spectral function
regularization \cite{EGM1} of the two-pion exchange contributions,
while the analysis by EM is based upon dimensionally regularized
expressions. Further differences can be attributed to the
implementation of the momentum-space cutoff in the Schroedinger
equation, the treatment of relativistic and isospin-breaking effects
as well as the fitting procedure:  the LECs, accompanying the
contact interactions, were determined by EM by fitting directly to
the scattering data, whereas EGM extracted them by the fit to the
Nijmegen PWA. For more details we refer the reader to the original
publications \cite{EGM,EM} and to the review articles \cite{EHM,ME}.

In its turn, the simultaneous extraction of the constants $c_D$ and
$c_E$ has recently been carried out in Ref.\,\cite{GQN}, where these
constants have been constrained by calculations of the binding
energies of the three-nucleon systems and of the reduced GT matrix
element of the triton $\beta$ decay. The nuclear wave functions were
generated in accurate {\it ab initio} calculations using both the
two-nucleon N$^3$LO EM \cite{EM} and three-nucleon N$^2$LO force
\cite{UVC,EEA}, whereas the reduced GT matrix element of the process
(\ref{3h}) has been calculated with the two-nucleon weak axial MECs,
derived from the same $\chi$PT Lagrangian \cite{GAZ} as the nuclear
forces. Besides, the calculations with two versions of the EGM
potential have also been performed (without, however, taking into
account the three-nucleon force). The resulting values of c$_D$
cover two rather distinct intervals (see Table \ref{tab:loh}).

In this work, we use the results of Ref.\,\cite{GQN} for $c_D$ to
calculate \loh. We apply the same potentials for generating the
deuteron and neutron-neutron ({\it{nn}}) wave functions and the weak
axial MEC operator, equivalent to that used in Ref.\cite{GQN}, thus
obtaining a rather consistent prediction for the values of \loh,
falling correspondingly into two rather different sets. Based on
these results we conclude that a precise measurement of \loh (i)
will provide a highly nontrivial consistency check for the chiral
EFT approach and (ii) using the extracted values of c$_D$ (\hdr),
will allow to calculate consistently other two-nucleon weak
processes, such as proton-proton fusion and solar neutrino
scattering on deuterons, which are important for astrophysics.
Besides it turns out \cite{GF} that \hdr enters also the capture
rate for the reaction
\mbox{$\pi^-\,+\,d\,\rightarrow\,\gamma\,+\,2\,n$}, which is the
best source of information on the {\it{nn}} scattering length
a$_{nn}$, see also a related work by Lensky et
al.~\cite{Lensky:2007zc}.

In addition, we employ the weak axial pion potential current
\cite{MRT} and study its influence on the values of \loh and \hdr as
well. As we shall see, its contribution to \loh is at the level of
the weak axial exchange charge density. Therefore, it should be
necessarily taken into account in the analysis of \loh. Since this
current was omitted in Ref.\,\cite{GQN}, its influence on extraction
of $c_D$ from the GT matrix element for the reaction (\ref{3h}) is
not clear.

Our manuscript is organized as follows. In Section \ref{prel}, we
discuss briefly the methods and inputs necessary for the
calculations, in Section \ref{rd}, we present the results and
discussion and we conclude in Section \ref{concl}. In \mbox{Appendix
\ref{appA}}, we give a short description of our treatment of the
{\it nn} wave functions within the K-matrix method and in Appendix
\ref{appB}, the form factors, arising after the Fourier
transformation of the weak MECs in presence of the regulator of the
type (\ref{REG}), are delivered.

\section{Methods and inputs}
\label{prel}

Here we present the necessary ingredients of the calculations of
\loh. They concern the used  potentials, the method of generation of
the {\it nn} wave functions, and the EFT currents.

\subsection{Nuclear potentials and wave functions}
\label{np}

In our calculations of nuclear matrix elements we start, in
accordance with Ref.\,\cite{GQN}, by adopting the N$^3$LO EM and two
species of the EGM($ORC$) potential with the parameters $O=2$, $R=0$ and
$C=4,5$. The meaning of these parameters is as follows: (i) $O=2$ means
that the potential is constructed at order N$^3$LO. (ii) The
value $R=0$ provides the potential for the usual non-relativistic
Lippmann-Schwinger equation. (iii) The value of $C=4(5)$ defines the
choice of the cutoff used in the Lippmann-Schwinger equation to be
$450$ MeV ($600$ MeV) and of the cutoff in the spectral function
representation of the two-pion potential to be $700$ MeV ($700$ MeV).
The {\it nn} wave functions in the continuum were generated by the
K(R)-matrix method \cite{GW,CDB}. The basic equations are detailed
in Appendix \ref{appA}.

For a reliable evaluation of \loh, the potential should describe
precisely the scattering length $a_{nn}$ and the effective range
$r_{nn}$ in the $^{1}$S$_0$ channel of the {\it nn} system. In Table
\ref{tab:annrnn},
\begin{table}[tb]
\caption{The scattering length a$_{nn}$ (in fm), and the effective
range r$_{nn}$ (in fm), calculated for the $^{1}$S$_0$ channel from
the used potentials.}
\begin{center}
\begin{tabular}{|l | c   c |  c  c    |}\hline
$^{1}$S$_0\,\,wave$          & a$_{nn}$ & r$_{nn}$ & a$^{exp}_{nn}$ &  r$^{exp}_{nn}$  \\\hline
EM       & -18.89  & 2.84 &  &   \\
EGM(204)   & -18.90  & 2.78 & -18.9 $\pm$ 0.4\,$^a$) & 2.75 $\pm$ 0.11\,$^b$)   \\
EGM(205)   & -18.91  & 2.88 &  &  \\\hline
\end{tabular}
\end{center}
$^a$) Ref.\,\cite{MS}\quad \quad $^b$) Ref.\,\cite{MNS}
\label{tab:annrnn}
\end{table}
we present these quantities for the used potentials, together with
the experimental values. We also present in Table \ref{tab:dwfs} the
properties of the used deuteron wave functions. Last but not least,
we emphasize that in the calculations of \loh, the {\it nn}
$^{2S+1}L_{j_f}$ partial waves with L=0,1,2,3 and j$_f$=0,1,2 are
taken into account.
\begin{table}[tb]
\caption{Deuteron properties derived from the EM and EGM(204) and
EGM(205) chiral potentials. Here E$_d$ is the deuteron binding
energy in MeV, P$_d$ is the D-state probability in \%, $<\mu>_z$ is
the deuteron magnetic moment in the nuclear Bohr magnetons, Q$_d$ is
the deuteron quadrupole moment in fm$^2$, $\sqrt<r^2>$ is the
deuteron root-mean-square radius in fm, A$_S$ is the asymptotic
normalization factor of the S-state in fm$^{-1/2}$ and $\eta_d$ is
the asymptotic D/S ratio. The magnetic and
quadrupole moments are calculated in the impulse approximation. \\
Experimental values of these quantities are: E$_d$=-2.224575(9) MeV
\cite{LA}, $<\mu>_z$=0.8574382284(94) \cite{MT}, Q$_d$=0.2859(3)
fm$^2$ \cite{ERC}, $\sqrt<r^2>$=1.9753(11) fm \cite{FMS},
A$_S$=0.8846(9) fm$^{-1/2}$ \cite{ERC}, and $\eta_d$=0.0256(4)
\cite{RK}.}
\begin{center}
\begin{tabular}{|l |  c  c  c  c  c  c  c |}\hline
 Deuteron    & E$_d$  & P$_d$ & $<\mu>_z$ &  Q$_d$ & $\sqrt<r^2>$ & A$_S$ & $\eta_d$  \\\hline
EM    & -2.224575  &4.52 &0.8540 &0.2753 &1.9750 &0.8843 &0.0256 \\
EGM(204)&-2.218923 &2.84 &0.8635 &0.2659 &1.9856 &0.8829 &0.0254 \\
EGM(205)&-2.223491 &3.63 &0.8590 &0.2692 &1.9754 &0.8833 &0.0255
\\\hline
\end{tabular}
\end{center}
\label{tab:dwfs}
\end{table}

\subsection{Weak nuclear MECs}
\label{wnceft}

The weak axial MECs constructed within the HB$\chi$PT \cite{PKMR}
and used in our calculations are discussed in detail
in Section 2.2 of Ref.\,\cite{RTMS}. We present in Table
\ref{tab:LECs} only the LECs necessary for the calculations of \loh.

The LECs $ c_i$ can be obtained from $\pi$N scattering, see
Ref.~\cite{VB} and references therein. There have been also attempts
to determine these LECs from two-nucleon scattering data
\cite{Rentmeester:1999vw}. Notice that NN N$^3$LO potentials of
Refs.~\cite{EGM,EM} adopt different values for the LEC $c_4$. In
particular, in  \cite{EM} the value of this LEC was tuned to improve
the description of the NN data. We further emphasize that the LECs
$c_{2,3,4}$ receive important contributions from the $\Delta$(1232)
isobar in the process of fixing them via saturation by higher
resonances, see Refs.~\cite{BKM1,Krebs:2007rh}.

\begin{table}[htb]
\caption{Values of the LECs c$_i$ (in GeV$^{-1}$) adopted in the
  N$^3$LO potentials
\cite{EM,EGM}.}
\begin{center}
\begin{tabular}{|l | c   c   c   c  c  |}\hline
Potential    & c$_1$  & c$_2$  & c$_3$ &  c$_4$ & c$_6$  \\\hline
EM           & -0.81  & 2.80 & -3.20 & 5.40 & 3.70 \\
EGM          & -0.81  & 3.28 & -3.40 & 3.40 & 3.70 \\\hline
\end{tabular}
\end{center}
\label{tab:LECs}
\end{table}
In the EM model, values of $c_4$ between $3.40$ and $5.40$ GeV$^{-1}$
are admissible \cite{EM}. As discussed above, the unknown LECs $\hat
d_1$ and $\hat d_2$, entering the weak axial MECs, are connected
with the LEC $c_D$ by Eq.\,(\ref{CED}).  In addition, we also
include the weak axial pion potential current of Eq.\,(A.17)
\cite{RTMS}, required by the PCAC constraint (\ref{NCEt}).

\begin{figure}[h!]
\centerline{ \epsfig{file=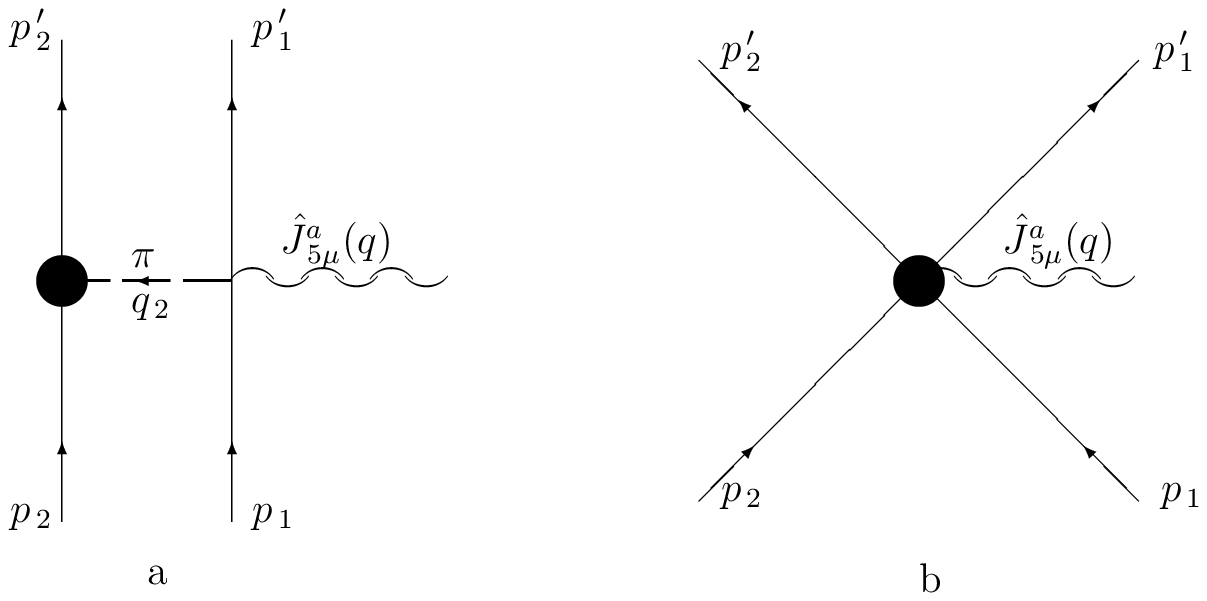} }
\vskip 0.4cm \caption{ The general structure of the two--nucleon
weak axial operators;
a-- the long-range operator, b-- the short-range operator. }

\label{figg1}
\end{figure}

The parameter $\hat d^R$ ($c_D$) manifests itself in the two-nucleon
contact vertex of the weak axial current, representing a short range
component of the MEC (see Fig.\,\ref{figg1}). Besides, it is also
present in a short-range part of the three-nucleon force. Fixing
this LEC from some few-nucleon processes  feasible in laboratory
allows one to make model independent predictions for other weak
processes (at the given order) triggered by this component of the
weak axial current. We further emphasize that the time component of
the weak axial MEC depends only on the known LECs $c_2$ and $c_3$ at
the same chiral order.

Besides these weak axial MECs, we include also the weak vector MECs
of the pion range: the $\pi$-pair term, the pion-in-flight term and
the $\Delta$ excitation current of the $\pi$ range, given in
Eqs.\,(A.9), (A.10) and (A.11), respectively \cite{RTMS}. Their
derivation is and properties are discussed in detail in Sections 2.1
and 2.2 of Ref.\,\cite{RTMS}. Let us note that the model independent
longitudinal $\pi$-pair and pion-in-flight terms, fixed by the low
energy theorem, satisfy the nuclear conserved vector current
constraint (\ref{NCVC}), whereas the current (A.11) is transverse
and, therefore, model dependent.

On the other hand, the leading order electromagnetic MECs were
obtained within the HB$\chi$PT approach for the $\pi N$ system in
Ref.\,\cite{PMR}, where they were referred to as "generalized tree
graphs". From these currents, the weak vector currents are derived
by rotating in the isospin space. They are given in Eqs.\,(35) and
(42) \cite{PMR}. It is straightforwardly seen that  our $\pi$-pair
and pion-in-flight currents coincide with the currents of Eq.\,(35)
and the $\Delta$ excitation current with the part of the current of
Eq.\,(42) proportional to the LEC $c_9$, saturated by the $\Delta$
isobar (see Eq.\,(43) \cite{PMR}) \footnote{Another part of the
current of Eq.\,(42) is proportional to the LEC $c^R_7$, which is
obtained by saturation by the $\omega$ meson. As it is seen from
Table 2 \cite{PMR}, the contribution from this current to the
reduced matrix element for the radiative neutron capture on proton
is minor and we do not consider it.}. Moreover, the same currents
were obtained in the Appendix C of the same Ref.\,\cite{PMR} from an
effective chiral Lagrangian of the $\pi N \Delta \rho \omega$ system
formulated in the heavy-fermion formalism. The most modern
derivation of the electromagnetic MECs \cite{Kolling:2009iq,KEKM},
based on the chiral effective theory of the $\pi N$ system and the
method of unitary transformation, again provides the leading order
currents of the above discussed form (see Eq.\,(4.28) \cite{KEKM}).
These currents contain some LECs, from which so far only  $\bar
d_{18}$ has been reliably fixed. The natural identification of the
leading order terms, proportional to $\bar d_{18}$, with the
standard $\pi$-pair and pion-in-flight currents leads to \be \bar
d_{18}=-1/4m^2_\pi=-12.8\, GeV^{-2}. \label{bd18} \ee On the other
hand, the value of  $\bar d_{18}$ was fixed in Ref.\,\cite{FM} by
the Goldberger-Treiman  discrepancy with the result \mbox{$\bar
d_{18}$=-10.14$\pm$0.45 $GeV^{-2}$}, which is in a good agreement
with the value (\ref{bd18}), predicted by the low energy theorem.
The other LECs, related to the transverse part of the discussed
current are not yet well known. Natural way to extract them seems to
be the study of pion photo- and electro-production, or capture
reactions. One can also proceed  by  fixing them via saturation by
the resonances, as it was done in Ref.\,\cite{PMR}.

It should be clear from our discussion that the leading order vector
MECs of the pion range are  determined by chiral invariance
uniquely. Therefore, working with these MECs in this approximation,
we can use safely our currents in the calculations done here.

In accord with Ref.\,\cite{GQN}, we multiply the MECs by the
regulator \be F^n_\Lambda(q^2)\,=\,e^{-(q/\Lambda)^{2n}}\,, \quad
n=2\,, \quad \Lambda=0.5\,\rm{GeV}\,,  \label{REG} \ee keeping the
currents local in configuration space. The form factors arising
after the Fourier transform of the currents are given in Appendix
\ref{appB}.

\section{Results and discussion}
\label{rd}

The calculation of \loh is discussed in detail in Section 3 of
Ref.\,\cite{RTMS}. Its value is obtained from the equation \be
\Lambda_{1/2}\,=\,\Lambda_{stat}\,+\,\delta\,\Lambda/3\,,
\label{loh} \ee where $\Lambda_{stat}$ and $\delta\,\Lambda$ are
given in Eqs.\,(3.2) and (3.7) \cite{RTMS}, respectively. The
results for \loh are displayed in Table \ref{tab:loh}.
\begin{table}[tb]
\caption{Calculated values of \loh, \bloh and of the MECs effect
$\delta$MECs=\loh-\loh (IA) (in s$^{-1}$) for the  chiral potentials
EM \cite{EM} and EGM \cite{EGM}. The capture rate \bloh differs from
\loh by taking into account the weak axial pion potential current.
The intervals of allowed values of c$_D$ are taken from
Refs.\,\cite{GQN,PNPC}, the corresponding values of
\hdr are derived using Eq.\,(\ref{reldrcd}) and c$_i$ from Table \ref{tab:LECs}.\\
EM$^{a}$ (EM$^{b}$)- the EM potential with c$_4$=3.4 (5.4) GeV$^{-1}$.}
\begin{center}
\begin{tabular}{|l |  c   c  | c   c |  c   c |  c   c  |}\hline
         &\quad\quad EM$^a$ &    &\quad\quad EM$^b$ &           & \quad\quad  EGM(204) &         & \quad\quad EGM(205) & \\\hline
$c_D$    & $-$0.05 & 0.2 & $-$0.3 & $-$0.1 & 0.05 & 0.33 & 3.30 & 4.20 \\
\hdr  & 1.18  & 1.44 & 2.23 & 2.44 & 1.28 & 1.57 & 4.66 & 5.60 \\
\loh    & 383.8  & 387.2  & 389.4 & 392.4 & 386.4 & 389.1 & 410.1 & 419.1      \\
\bloh   & 381.6  & 385.1 & 387.2 & 390.1 & 385.2 & 387.9 & 408.3 & 417.1 \\
$\delta$MECs  & 4.2  & 7.6 & 9.8 & 12.4 & 2.1 & 4.7 & 29.9 & 38.5
\\\hline
\end{tabular}
\end{center}
\label{tab:loh}
\end{table}
As it is seen from Table \ref{tab:loh}, the values of \loh are
located in two rather distinct intervals. For the NN potentials EM
and EGM(204), they are in the interval 383.8 s$^{-1}$ $\le$ \loh
$\le$ 392.4 s$^{-1}$, thus featuring a spread of about  $2\%$. The
spread of the values of \loh obtained in the EGM(205) model is about
the same, but these values are larger by $\approx$ 6 \% and the MECs
effect is also much larger. If one would like to get within the
model EGM(205) the value \loh=389.1 s$^{-1}$, one should take
\hdr=2.35 corresponding to $c_D$=1.08 which is far out of the
interval 3.30\,$\le$\,$c_D$\,$\le$\,4.20 allowed by the reduced GT
matrix element of the triton $\beta$ decay.

One can see from Table \ref{tab:loh} that the presence of the weak
axial pion potential current suppresses \loh up to 0.5 \%. E.g., in
the case of the model EM$^b$ and for the value \hdr (c$_D$) =2.23
(-0.3), one gets \loh=389.4 s$^{-1}$, whereas \bloh=387.2 s$^{-1}$.
This difference in the capture rate can be reflected in the change
of \hdr (c$_D$) = 2.07 (-0.45), which means the change of $\sim$ 7.5
\% (44 \%) in \hdr (c$_D$).

Let us now discuss for the same model EM$^b$ the variation of
extracted \hdr (c$_D$), if the supposed experimental value of
\loh=390 $\pm$6 s$^{-1}$ is measured with an  accuracy of 1.5 \%.
The extracted value of \hdr=2.27$\pm$0.44, to which corresponds
c$_D$=-0.26$\pm$0.42. It is seen that the value of \hdr (c$_D$)
would be known with an accuracy of 20 \% (140 \%). With the weak
axial pion potential current included, the interval of \hdr (c$_D$)
would be shifted to the negative values by  $\sim$ 7.5 \% (44 \%) as
a whole.

In Table \ref{tab:pacon}, we give individual contributions of
various components of the currents, the multipoles of which can be
found in Appendix C of Ref.\,\cite{RTMS}. For the EGM version, the
contribution of the weak axial pion potential current $\vec
{\jmath}_A$(p.p.c.) is about the same as that of the time component
of the weak axial MECs, and for the EM potentials it is twice as
large. Notice that in the calculations of \cite{RTMS} based on the
Nijmegen I potential \cite{SKTS},  the suppression of \loh by $\sim$
1 \% was obtained for this current, whereas in Ref.\,\cite{PI1} no
effect has been found based on the nuclear wave functions generated
by the Argonne $v_{18}$ potential \cite{WSS}. For the EGM(205)
version, the contribution of the short-range  weak axial MECs, $\vec
{\jmath}_A$(EFT), is strongly enhanced as a consequence of the large
value of \hdr.
\begin{table}[b]
\caption{Contributions of various components of the currents to \loh
(in s$^{-1}$) obtained using  chiral potentials EM \cite{EM} and EGM
\cite{EGM} for certain values of \hdr. IA($^{1}S_0$)- the
contribution to \loh from the $^{1}S_0$ {\it nn} final state and the
one-body currents; + ${\jmath}^0_A$- with the time component of the
weak axial MECs added; +$\vec {\jmath}_A$(EFT)- with the space
component of the EFT weak axial MECs \cite{PKMR} added; +$\vec
{\jmath}_V$- with the space component of the weak vector MECs added;
IA(full)- the contribution to \loh from all the {\it nn} partial
waves taken into account and the one-body currents; + ${\jmath}_A$-
with the weak axial MECs \cite{PKMR} added; +$\vec {\jmath}_V$- with
the space component of the weak vector MECs added; +$\vec
{\jmath}_A$(p.p.c.) - with the weak axial pion potential current
added. The values of the doublet capture rate in the last column
present in fact \bloh. }
\begin{center}
\begin{tabular}{|l |  c  |  c   c   c   c  | c   c   c | c |}\hline
Potential & \hdr&IA($^{1}S_0)$&+${\jmath}^0_A$&+$\vec
{\jmath}_A$(EFT)&+$\vec {\jmath}_V$&IA(full)&+${\jmath}_A$&+$\vec
{\jmath}_V$&+$\vec{\jmath}_A$(p.p.c.) \\\hline
EGM(204) &  1.57  & 245.2  & 244.1  & 248.3 & 249.8 & 384.3 & 387.3 & 389.1 & 387.9   \\
EGM(205) &  5.13  & 241.8  & 240.7  & 274.7 & 276.6 & 379.5 & 412.4 & 414.6 & 412.7   \\
EM$^{a}$ &  1.18  & 240.9  & 240.0  & 243.0 & 244.9 & 379.6 & 381.6 & 383.8 & 381.6   \\
EM$^{b}$ &  2.23  & 240.9  & 240.0  & 248.5 & 250.5 & 379.6 & 387.2
& 389.4 & 387.7   \\\hline
\end{tabular}
\end{center}
\label{tab:pacon}
\end{table}

In Table \ref{tab:comp1}, we compare the results of our IA
calculations for the model EM$^{b}$ with the second  row of Table II
of Ref.\,\cite{PI3}. As it is seen, the agreement between the
calculations is very good. On the other hand, the comparison of the
final result \loh=392.0$\pm$2.3 s$^{-1}$ \cite{PI2}  with our Table
\ref{tab:loh} shows reasonable agreement in calculations for the
potential model EM$^{b}$, whereas the result \loh=399$\pm$3 s$^{-1}$
\cite{PI3} indicates the difference of $\sim$ 2 - 3 \%. In general,
our results exhibit much more model dependence of \loh, than
admitted in Refs.\,\cite{PI2,PI3}.
\begin{table}[tb]
\caption{Contributions of the {\it nn} partial waves to  \loh (in
s$^{-1}$), calculated for the model EM$^{b}$ and with the IA
currents. The third row contains the second row of Table II
\cite{PI3}.   }
\begin{center}
\begin{tabular}{|l |  c   c   c   c   c   c   c   |}\hline
         & $\delta{\Lambda}_{1/2}(^{1}S_0$)&$\delta{\Lambda}_{1/2}(^{3}P_0$)  &$\delta{\Lambda}_{1/2}(^{3}P_1$) &
         $\delta{\Lambda}_{1/2}(^{3}P_2$) & $\delta{\Lambda}_{1/2}(^{1}D_2$) & $\delta{\Lambda}_{1/2}(^{3}F_2$)
         & ${\Lambda}_{1/2}$ \\\hline
This work         & 240.9 & 22.0 & 40.9 & 69.4 & 6.0 & 0.4 & 379.6 \\
Ref.\,\cite{PI3} & 238.8 & 21.1 & 44.0 & 72.4 & 4.5 & 0.9 & 381.7
\\\hline
\end{tabular}
\end{center}
\label{tab:comp1}
\end{table}

Let us note that our results for \loh contain the factor 1.028(4)
taking into account the radiative corrections. These corrections
have not yet been calculated within the EFT. The result of
"classical" calculations \cite{CMS} for the inner correction is
0.024(4), valid for the muon capture in any nucleus. The vacuum
polarization correction to the muon bound state wave function must
be evaluated for every nucleus separately. For the muon capture in
deuteron, this correction is $1.80\,\alpha/\pi$=0.0042 \cite{CPC}.
Then the overall radiative corrections for the muon capture in
deuteron turn out to be the same as for the muon capture on proton.

\section{Conclusions}
\label{concl}

In this work, we studied the doublet capture rate, \loh, for the
reaction $\mu^-(d, \, 2n)\nu_\mu$ (\ref{mudc}) within the framework
of chiral EFT. The nuclear wave functions of the initial and final
states were generated from the N$^{3}$LO  chiral potentials of Entem
and Machleidt \cite{EM} and of  Epelbaum, Gl\"ockle and Mei{\ss}ner
\cite{EGM}. The properties of the $^{1}S_0$ {\it nn} state and of
the deuteron are presented in Table \ref{tab:annrnn} and Table
\ref{tab:dwfs}, respectively. The employed weak axial MECs were also
constructed within the same EFT scheme \cite{PKMR,DG}. The
corresponding LECs, taken consistently with the potentials, are
listed in  Table \ref{tab:LECs}. The short-range part of the weak
axial MECs contains one unknown LEC, \hdr ($c_D$), which has been
constrained earlier \cite{GQN} by Gazit, Quaglioni and Navr\'atil in
the study of the GT matrix element of the triton $\beta$ decay and
of the static properties of the 3N system, using the same NN
potentials and weak axial MECs. With the constant \hdr, constrained
in this way and with the same regulator (\ref{REG}) entering the
weak MECs, we performed rather consistent calculations of \loh. As
it is seen from Table \ref{tab:loh}, the results are located in two
rather distinct intervals, in which the spread of \loh is about 2
\%. It follows from this result that (i) studying only the triton
$\beta$ decay or the muon capture in deuterium might be insufficient
for a precise determination of \hdr , (ii) combined investigation of
both reactions can not only provide precise value of \hdr, but will
allow for a highly nontrivial test of the chiral EFT framework.

In addition, we studied the influence of the weak axial pion
potential current on the \loh and \hdr. This current is not a part
of the weak axial MECs derived within the HB$\chi$PT in
Refs.\,\cite{PKMR,DG}, and was not taken into account in the
calculations \cite{GQN} either. As it is seen from our Table
\ref{tab:loh} and Table \ref{tab:pacon}, its contribution to \loh is
non-negligible and can change the value of \hdr (c$_D$) up to 7.5 \%
(44 \%). In our opinion, the full consistency of calculations cannot
be achieved without including this current also in the study of the
reaction (\ref{3h}).

What are the possible origins of the observed fairly large spread in
the calculated values of \hdr when using the EGM NN potential of
Ref.~\cite{EGM}? First of all,  the value of the LEC $c_D$ was
obtained in Ref.~\cite{GQN} without taking into account the 3NF in
the case of the EGM potentials. Given the significant underbinding
of the triton for N$^3$LO NN potentials of Ref.~\cite{EGM} without
the inclusion of the 3NF \footnote{The underbinding is of the order
of $1\ldots 1.5$ MeV
  which is larger than in the case of the N$^2$LO potentials.}, the
obtained values for $c_D$ in the case of the EGM NN potentials
should be taken with care. Furthermore, while the applied regulator
keeps the calculations in the configuration space local, the
regulator adopted in the N$^3$LO NN potentials  \cite{EM,EGM} reads
\be F_\Lambda(p',p)\,=\,e^{-p'^6/\Lambda^6-p^6/\Lambda^6}\,, \quad
\Lambda=0.5\,\rm{GeV}\,,  \label{REGnl} \ee where $p$ ($p'$) is the
initial (final) nucleon momentum in the center-of-mass system.
Clearly, this regulator leads to  strong non-localities in
configuration space. Strictly speaking, only calculations with this
type of the regulator can be considered as fully consistent. It is
desirable to carry out such calculations in the future in order to
see how the form of the regulator affects the results for \loh and
the values of extracted constant \hdr ($c_D$). Another possible
source of inconsistency emerges from using two- and three-nucleon
forces and the exchange currents at different orders in the chiral
expansion. The N$^3$LO NN potentials should be accompanied with the
3NFs calculated at the same order in the chiral expansion. Only
recently, the N$^3$LO corrections to the 3NF have become available
\cite{Ishikawa:2007zz,Bernard:2007sp,Bernard:2011zr}. Moreover, the
results of the pioneering work \cite{Skibinski:2011vi}  indicate
that the inclusion of the long-range N$^3$LO corrections in the 3NF
might have strong impact on the values of the LECs $c_D$ and $c_E$.
This issue needs to be investigated in the future. Finally, we
applied the weak vector MECs in the leading order only. Evidently,
the estimation of higher order terms, derived in Refs.\,
\cite{PMR,Kolling:2009iq,KEKM,Pastore:2009is,PGSV}, is mandatory in
future calculations. Last but not least, it would be interesting to
carry out the calculations using the N$^2$LO NN potentials and the
corresponding 3NF. This would help to obtain a reliable estimation
of the theoretical uncertainty for \loh.

\section*{Acknowledgments}
We thank P. Navr\'atil, P. Kammel and L.E. Marcucci for discussions
and correspondence. The correspondence with A. Czarnecki is
acknowledged. The research by M.T. was partially supported by the
Czech Ministry of Education, Youth and Sports within the project
LC06002 and by the grant GA \v{C}R P203/11/0701, and the work of
R.M. was supported in part by the US Department of Energy under
Grant No. DE-FG02-03ER41270. E.E.~acknowledges the support by the
European Research Council (ERC-2010-StG 259218 NuclearEFT).

\newpage

\appendix

\section{The {\it nn} wave functions}
\label{appA}

Here we discuss in more detail the derivation of the {\it nn} wave
functions in the formalism of the K-matrix, referring essentially to
Ch.\,6 and Ch.\,7 of Ref.\,\cite{GW} and to Appendix A of
Ref.\,\cite{CDB}.

The basic equation for the {\it nn} scattering wave function in
terms of the T-matrix is
\be
\Psi^+_{\vec \kappa,S\nu}(\vec r)\,=\,\chi_{\vec \kappa,S\nu}(\vec r)+\sum_{S'\nu'}\,
\int\,\frac{d^3{\vec\kappa}'\chi_{{\vec\kappa}',S'\nu'}}{\epsilon(\kappa)+i\eta-\epsilon(\kappa')}\,
\left<{\vec \kappa}',S'\nu'|T|\vec \kappa,S\nu\right> \, ,   \label{be}
\ee
where the plane wave is defined as
\be
\chi_{\vec \kappa,S\nu}(\vec r)\,=\,e^{i\vec \kappa\cdot \vec r}\,u_{S\nu}\,,  \label{pw}
\ee
further
\be
\epsilon(\kappa)\,=\,\kappa^2/M_n\,,   \label{ek}
\ee
and $\vec \kappa$  (M$_n$) is the relative {\it nn} momentum (neutron mass).

We expand the scattered wave as
\be
\Psi^+_{\vec \kappa,S\nu}(\vec r)\,=\,4\pi\sum_{lJ\nu' l' S'}\left<\hat r,l' S'\nu'|{\cal J}^J|
\hat \kappa,lS\nu\right>\,\Psi^+_{l'S',\kappa JlS}(r)u_{S'\nu'}\,, \label{Psif}
\ee
and the T-matrix as
\be
\left<{\vec \kappa}',S'\nu'|T|\vec \kappa,S\nu\right>\,=
\,\sum_{ll'J}\,\left<{\hat \kappa}',l' S'\nu'|{\cal J}^J|
\hat \kappa,lS\nu\right>\,T^J_{l'S',lS}(\kappa',\kappa)\,.  \label{Tf}
\ee
Then introducing the definitions
\bea
\Psi^+_{l',lJ}(r)\,&\equiv&\,\Psi^+_{l'1,\kappa Jl1}(\kappa,r)\,,   \label{pwfc} \\
T^J_{l',l}(\kappa',\kappa)\,&\equiv&\,T^J_{l'1,l1}(\kappa',\kappa)\,,  \label{pwt}
\eea
we obtain the equation,
\be
\Psi^+_{l',lJ}(r)\,=\,i^{l'}\left\{\delta_{l',l}\,j_l(\kappa r)
+ \int\,\frac{{\kappa'}^2 d\kappa' j_{l'}(\kappa' r)}
{\epsilon(\kappa)+i\eta-\epsilon(\kappa')}\,T^J_{l',l}(\kappa',\kappa)\right\}\,. \label{bepw}
\ee
Let us write the function $\Psi^+_{l',lJ}(r)$ as
\be
\Psi^+_{l',lJ}(r)\,=\,i^{l'}\,\frac{w^J_{l',l}(\kappa,r)}{\kappa r}\,.  \label{dw}
\ee
Then  from Eq.\,(\ref{bepw}), one obtains the following equation for
the function $w^J_{l',l}(\kappa,r)$
\be
\frac{w^J_{l',l}(\kappa,r)}{\kappa r}\,=\,\delta_{l',l}\,j_l(\kappa r) + \int\,\frac{{\kappa'}^2 d\kappa' j_{l'}(\kappa' r)}
{\epsilon(\kappa)+i\eta-\epsilon(\kappa')}\,T^J_{l',l}(\kappa',\kappa)\,.   \label{eqfw}
\ee
It can be shown that the new functions, defined as
\be
{\bar w}^J_{l',\alpha}(\kappa,r)\,=\,e^{-i\delta^J_\alpha}\,\sum_{l}\,
w^J_{l',l}(\kappa,r)U^J_{l1,\alpha}\,,
\label{bw}
\ee
satisfy the equations,
\be
\frac{\bar w^J_{l',\alpha}(\kappa,r)}{\kappa r}\,=
\,e^{-i\delta^J_\alpha}\left\{\,j_{l'}(\kappa
r)U^J_{l'1,\alpha}
 + \int\,\frac{{\kappa'}^2 d\kappa' j_{l'}(\kappa' r)}
{\epsilon(\kappa)+i\eta-\epsilon(\kappa')}\,\sum_l\,T^J_{l',l}
(\kappa',\kappa)U^J_{l1,\alpha}\right\}\,.   \label{eqfbw}
\ee
The unitary matrix $U^J_{l1,\alpha}$ is defined as
\be
U^J_{l1,\alpha}\,=
\,\left( \begin{array}{cc} \cos\,\epsilon^J & -\sin\,\epsilon^J \\
                           \sin\,\epsilon^J &  \cos\,\epsilon^J
                                   \end{array} \right)\,,  \label{UJ}
\ee
where $\epsilon^J$ is the mixing angle.
The relation between the T- and K-matrices reads
\bea
T^J_{l',l}(\kappa',\kappa)\,&=&\,K^J_{l',l}(\kappa',\kappa)-i\pi M_n \kappa\,\sum_{l''}\,
K^J_{l',l''}(\kappa',\kappa)T^J_{l'',l}(\kappa,\kappa)   \nonumber \\
&=&\,\sum_{l''}\,K^J_{l',l''}(\kappa',\kappa)[\delta_{l'' l}-i\pi
M_n \kappa\,\sum_{\alpha}\, U^J_{l'' 1,\alpha} e^{i\delta^J_\alpha}
\cos \delta^J_\alpha K^J_{\alpha}(\kappa) U^J_{\alpha,l 1}]\,.
\label{TKR}
\eea
Here
\be
K^J_{\alpha}(\kappa)\,\equiv\,K^J_{\alpha}(\kappa,\kappa)\,=
\,-\frac{1}{\pi M_n \kappa}tg \delta^J_\alpha\,.  \label{KONSH}
\ee
Using Eq.\,(\ref{TKR}) in Eq\,(\ref{eqfbw}), one obtains for the functions
$\bar w_{l',J\alpha}(\kappa,r)$ the new set of equations in terms of
the K-matrix only
\bea
\frac{\bar w^J_{l',\alpha}(\kappa,r)}{\kappa r}\,&=&
\,\cos \delta^J_\alpha \left\{j_{l'}(\kappa r)U^J_{l'1,\alpha}
 + \int\,\frac{ d\kappa' }{\epsilon(\kappa)-\epsilon(\kappa')}
 \left[{\kappa'}^2 j_{l'}(\kappa' r)\,
\sum_l\,K^J_{l',l}(\kappa',\kappa)U^J_{l1,\alpha} \right.\right.\nonumber \\
&&\left.\left.\,-\kappa^2 j_{l'}(\kappa r) K^J_{\alpha}(\kappa) U^J_{l' 1,\alpha}\right]\right\}\,.
\label{eqfbwf}
\eea
It holds that
\be
K^J_{\alpha}(\kappa)U^J_{l' 1,\alpha}\,=\,\sum_l\,K^J_{l',l}(\kappa,\kappa)U^J_{l1,\alpha}\,.
\label{EQFK}
\ee
The functions $\bar w^J_{l',\alpha}(\kappa,r)$ were
computed from Eqs.\,(\ref{eqfbwf}) by using the program phases
\cite{RMP}. Resolving Eq.\,(\ref{bw}) for the functions
$w^J_{l',l}(\kappa,r)$, one obtains
\be
w^J_{l',l}(\kappa,r)\,=\,\sum_\alpha\,e^{i\delta^J_\alpha}\,
{\tilde U}^J_{\alpha,l1}\,\bar w^J_{l',\alpha}(\kappa,r)\,. \label{req} \ee
Analogously, for the non-coupled channels we have
\be
\Psi^+_l(\kappa,r)\,=\,i^l e^{i\delta_l}
\frac{w_l(\kappa,r)}{\kappa r}\,,  \label{ncc}
\ee
where the function $w_l(\kappa,r)$ satisfies
the equation,
\bea
\frac{ w_{l}(\kappa,r)}{\kappa r}\,&=&\,\cos \delta_l
\left\{j_{l}(\kappa r) + \int\,\frac{ d\kappa' }
{\epsilon(\kappa)-\epsilon(\kappa')}\left[{\kappa'}^2 j_{l'}(\kappa' r)\,
K_{l}(\kappa',\kappa) \right.\right.\nonumber \\
&&\left.\left.\,-\kappa^2 j_{l}
(\kappa r)K_{l}(\kappa)\right]\right\}\,. \label{eqfwncc}
\eea

\section{The form factors arising after the Fourier transformation
of the weak MECs due to the regulator (\ref{REG})}
\label{appB}

Here
we present the form factors arising due to the regulator (\ref{REG})
after the Fourier transformation of the weak MECs containing one
boson propagator with the mass m$_B$.
\be
W^n_{0B}\,=\,\frac{2}{\pi x_B}\,I^n_1(x_B,\bar\Lambda)\,, \label{W0B}
\ee
\be
W^n_{1B}\,=\,\frac{2}{\pi x^2_B}\,
\left[I^n_1(x_B,\bar\Lambda)-I^n_2(x_B,\bar\Lambda)+x_B\,I^n_3(x_B,\bar\Lambda)\right]\,,
\label{W1B}
\ee
\be W^n_B\,=\,\frac{2}{\pi x^3_B}\,I^n_4(x_B,\bar\Lambda)\,,  \label{WB}
\ee
\be
W^n_{2B}\,=\,W^n_{0B}\,+\,\frac{3}{x_B}\,W^n_{1B}\,-\,W^n_B\,, \label{W2B}
\ee
where
\be
I^n_1(x_B,\bar\Lambda)\,=\,\int^\infty_0\,dt\
\frac{t\, \sin (t x_B)}{1+t^2}\ e^{(-t/\bar\Lambda)^{2n})}\,, \label{I1}
\ee
\be
I^n_2(x_B,\bar\Lambda)\,=\,\int^\infty_0\,dt\
\cos (t x_B)\ e^{(-t/\bar\Lambda)^{2n})}\,, \label{I2}
\ee
\be
I^n_3(x_B,\bar\Lambda)\,=\,\int^\infty_0\,dt\
\frac{ \cos (t x_B)}{1+t^2}\ e^{(-t/\bar\Lambda)^{2n})}\,, \label{I3}
\ee
\be
I^n_4(x_B,\bar\Lambda)\,=\,\int^\infty_0\,dt\
\sin (t x_B)\,e^{(-t/\bar\Lambda)^{2n})}\,, \label{I4} \ee
\be x_B\,=\,m_B r\,,\quad \bar\Lambda\,=\,\Lambda/m_B\,.    \label{xBbL}
\ee
For n=1, Eqs.\,(A.24)-(A.28) of Section A.2 of Ref.\,\cite{RTMS} are
reproduced.

\end{document}